\documentclass{jfm}
\usepackage{graphicx}
\usepackage[export]{adjustbox}
\usepackage{color}

\shorttitle{Self-similarity in the breakup of very dilute viscoelastic solutions}
\shortauthor{A. Deblais, M. A. Herrada, J. Eggers and D. Bonn}

\title{Self-similarity in the breakup of very dilute viscoelastic solutions}

\author{A. Deblais\aff{1}
  \corresp{\email{A.Deblais@uva.nl}},
  M. A. Herrada\aff{2}, J. Eggers\aff{3} and D. Bonn\aff{1}}

\affiliation{\aff{1}Van der Waals-Zeeman Institute, Institute of Physics, University of Amsterdam, 1098XH Amsterdam, The Netherlands.
\aff{2}Depto. de Mecánica de Fluidos e Ingeniería Aeroespacial, Universidad de Sevilla, E-41092 Sevilla, Spain.
\aff{3}School of Mathematics, University of Bristol, University Walk, Bristol BS8 1TW, UK.}

\begin{document}

\maketitle

\begin{abstract}
When pushed out of a syringe, polymer solutions form droplets attached by long and slender cylindrical filaments whose diameter decreases exponentially with time before eventually breaking. In the last stages of this process, a striking feature is the self-similarity of the solution shape near the end of the filament. This means that shapes at different times, if properly rescaled, collapse onto one universal shape. A theoretical description inspired by this similarity observation and based on the Oldroyd-B model was recently shown to disagree with existing experimental results.
By revisiting these measurements and analysing the interface profiles of very diluted polyethylene oxide solutions at high temporal and spatial resolution, we show that they are very well described by the model. 
\end{abstract}


\section{Introduction}

The formation of drops has become a paradigm for the study of singularities in fluid mechanics and beyond. The formation of a drop from an orifice leads to a new length scale, the diameter of the neck that connects the drop to the orifice, that goes to zero at a finite time when the drop breaks off from the orifice. As a result, pinch-off is described by a similarity solution that describes the time evolution and self-similar shape of the neck close to breakup. In the case of Newtonian fluids of both large and small viscosity, the neck diameter behaves like a power law as a function of the time to pinch-off. The interface is found to have a universal shape, and profiles at different times can be superimposed onto one another by rescaling the radial and axial coordinates by appropriate powers of the time distance to the singularity.

\begin{figure}
  \centerline{\includegraphics[scale=0.45]{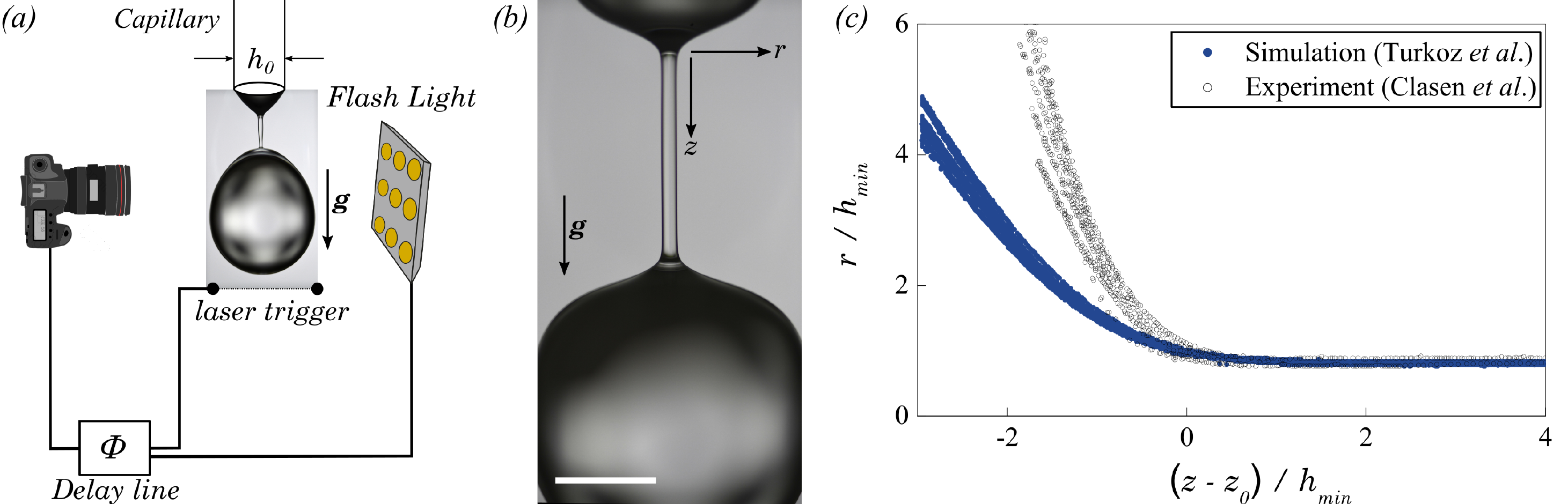}}
  \caption{(a) Schematic of the experimental setup used to determine the interface shape of the polymer thread. A full-frame camera is combined with a flash light allowing short flash duration. The flash is triggered on the falling drop and a high accuracy delay line allows to shift the breakup event in time. (b) Typical photograph of a pendant drop of a very dilute polymer solution breaking from a syringe. A long tiny polymer thread connecting the two drops is formed. Scale bar is 1mm. (c) Self-similar thinning of the interface profiles obtained from the experimental work of \cite{Clasen2006} and the simulations of \cite{Turkoz2018}. $z_{0}$ is the axial location for which the profiles collapse best. The comparison reveals a discrepancy between the two. Figure adapted from \cite{Turkoz2018}.}
\label{fig:FigThreadTurkoz}
\end{figure}

Beyond Newtonian fluids, much work has been dedicated to the formation of drops in non-Newtonian fluids such as polymer solutions. These fluids are characterized by a slow time scale $\lambda$ on which the constituents relax. If one takes a dilute solution of a high-molecular weight polymer and tries to make a droplet \citep{Middleman1965,Goldin1969,Petrie1976,Eggers1997,Deblais2018b}, polymers become stretched in the extensional flow close to pinch-off, and long and slender filaments form in between drops, where previously power-law pinch-off would have been observed \citep{Goldin1969,Bazilevskii1981,Entov1984,Wagner2005,Bhat2010,Clasen2006}. In the case of a jet, this means a long sequence of almost circular drops form, connected by tiny threads, a phenomenon which has been called the ``beads-on-a-string'' structure.
In the case of a dripping faucet, one observes essentially the same phenomenon, except that here only a single drop forms, 
with perhaps a satellite drop in between \citep{Wagner2005}, connected to the faucet by a tiny thread, as seen in Fig.~\ref{fig:FigThreadTurkoz}(a,b). 

This and similar phenomena are often modelled using the so-called Oldroyd-B model \citep{Bird1987}, which describes the polymer relaxation with a single time scale $\lambda$. The stress is written as the sum of a polymeric contribution and that of the solvent. 

 The formation of filaments and their instabilities have by now become a benchmark problem for testing viscoelastic fluid mechanics \citep{Anna2001,Mckinley2002,Furbank2004,Suryo2006,Smith2010,Miskin2012,Huisman2012}.
Instead of following a power law, the filament now thins exponentially with a rate set by the relaxation time of the polymer \citep{Bazilevskii1981,Anna2001,Clasen2006}. The profile is extremely uniform over the thread, and then merges smoothly with a neighboring drop on either side. It was proposed by \cite{Clasen2006} that the profile at this junction is once more a similarity solution, which yields a universal profile, if both the axial and the radial coordinate is rescaled with the thread radius, with an exponential rise toward the drop (Fig.~\ref{fig:FigThreadTurkoz}(c)). \cite{Clasen2006} were able to calculate the similarity profile using a lubrication approximation, in which the interface slope is assumed small; however, this assumption of small slopes is not satisfied throughout the profile. Indeed, while an experiment using long flexible polymers in a viscous solvent showed collapse to a self-similar profile, a comparison with the theoretical calculation revealed an axial length scale in the experiment that was about twice as short as the one obtained from lubrication theory (Fig.~\ref{fig:FigThreadTurkoz}(c)). At the time, this serious discrepancy in the self-similar profile was attributed to a failure of the lubrication approximation. 

However, a recent full numerical simulation of the Oldroyd-B equations \citep{Turkoz2018} showed the same discrepancy with the experimental data of \cite{Clasen2006}, and rather agreed with the lubrication calculation (Fig.~\ref{fig:FigThreadTurkoz}(c)). This left the serious possibility that the source of the discrepancy was the Oldroyd-B model itself. One possibility was that the experimental fluid was described by more than a single length scale, the other that the finite extensibility of a real polymer has to be taken into account, as described for example by more elaborate models such as the FENE-P model \citep{Bird1987}, which also describes the shear-thinning behavior of a real polymeric fluid. 

\begin{figure}
  \centerline{\includegraphics[scale=0.95]{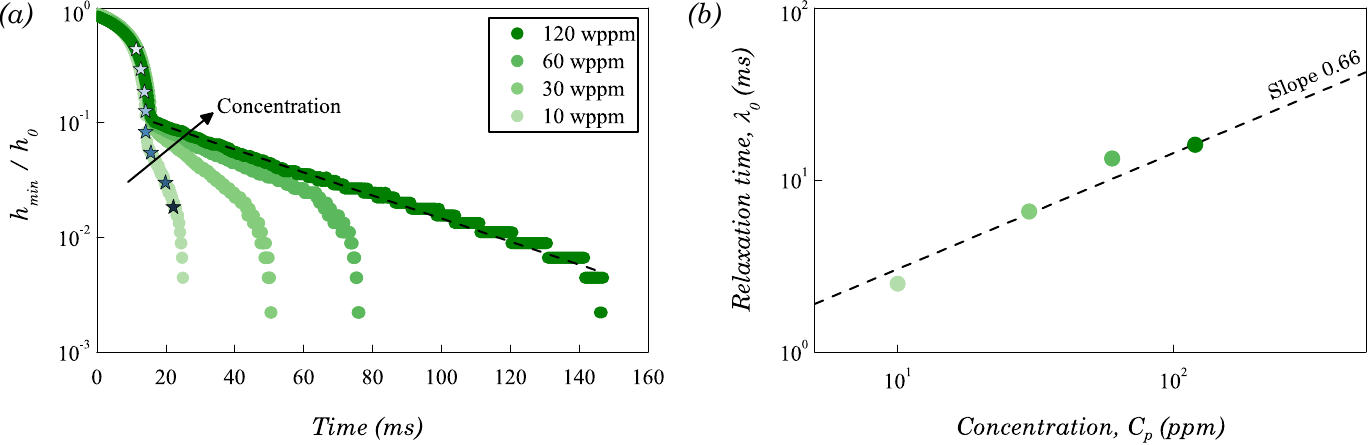}}
  \caption{Thinning dynamics of a filament of four concentrations of polyethylene oxide in water. (a) Symbol colours, from light to dark: 10 wppm, 30 wppm, 60 wppm, and 120 wppm. The minimum neck radius $h_{min}$ is tracked in time and normalized by the inner radius $h_{0}$ of the syringe orifice; the longest relaxation time of the solution $\lambda_{0}$ is deduced from the slope of the elasto-capillary regime highlighted by the dashed black line. (b) Relaxation time $\lambda_{0}$ as function of the concentration $c_{p}$. The dashed line is a power law fit to the experimental points $\lambda_{0}\propto {C_{p}}^{0.66}$.}
\label{fig:ThinningDynamics}
\end{figure}

More recently, the similarity theory of \cite{Clasen2006} 
was extended to a treatment of the full axisymmetric Oldroyd-B equations, and the similarity profile was calculated without any lubrication assumptions \citep{Snoeijer2019}. The robustness of these calculations was also underlined by the observation that the universal interface shape even holds for purely elastic filaments undergoing elastocapillary instabilities \citep{Snoeijer2019, Eggers2020} that could be dubbed sausage-on-a-string instabilities \citep{Mora2010}. These findings suggest that the specific type of viscoelastic model for the polymer solution is not crucial for calculating the shape of the interface, and hence could not explain the discrepancy between theory and experiment. 

To clear up these questions, we investigate the breakup of very dilute polymer solutions. We record the interface profiles of the filaments at an extremely high temporal and spatial resolution during the experiments, and compare the results with the newly developed similarity theory of \cite{Eggers2020}, based on the full Oldroyd-B equations. Our results convincingly show that the experimental profiles all converge to a universal self-similar solution that in addition agrees excellently with theory. 

\section{Experiments}

We experimentally study the extensional thinning and destabilization of filaments of long-chain polymer solutions in water at different concentrations (Fig. \ref{fig:ThinningDynamics}). The experiments are performed with polyethylene oxide (PEO) of a molecular weight ($M_{w}$) of $\rm 4 \pm 2 \times 10^{6}~g \ mol^{-1}$ \citep{Berman1978} from Sigma-Aldrich (purity=0.98). Four concentrations $C_{p}$ between 10 and 120 wppm are obtained from successive dilutions of the initial batch. This range of concentrations is chosen to be well below the critical overlap concentration of polymer coils \citep{Graessley1980}:

\begin{equation}
    c^{*} = \frac{0.77}{[\eta]} ,
    \label{eq:cstar}
\end{equation}

with $[\eta]$ the intrinsic viscosity of the polymer solution (See details in Table \ref{tab:FluidParameters}, \cite{DelGiudice2017}). The concentrations we consider allow us to work in conditions that are very close to satisfying the Oldroyd-B model, as we will show below.  A syringe pump supplies the polymer solutions to the needle tip (inner diameter $h_{0}$=2 mm). 

\begin{table}
  \begin{center}
\def~{\hphantom{0}}
  \begin{tabular}{lcccc}
      $C_{p} (wppm)$   & [$\eta$]/($m^{3}/kg$) & $c^{*}$ & $C_{p}/c^{*}$ & $n$/($10^{18} m^{-3}$)\\[3pt]
       10 &  1.408 & 0.071 & 0.014 & $\rm 1.51$\\
       30 &  '' & '' & 0.042 & $\rm 4.52$\\
       60 &  '' & '' & 0.084 & $\rm 9.03$\\
       120 & '' & '' & 0.169 & $\rm 18.1$\\
  \end{tabular}
  \caption{Physical parameters of the polymer solutions investigated (PEO) with a molecular weight $M_{w} = \rm 4 \times 10^{6}~g \ mol^{-1}$.}
  \label{tab:FluidParameters}
  \end{center}
\end{table}

\begin{figure}
  \centerline{\includegraphics[scale=0.6]{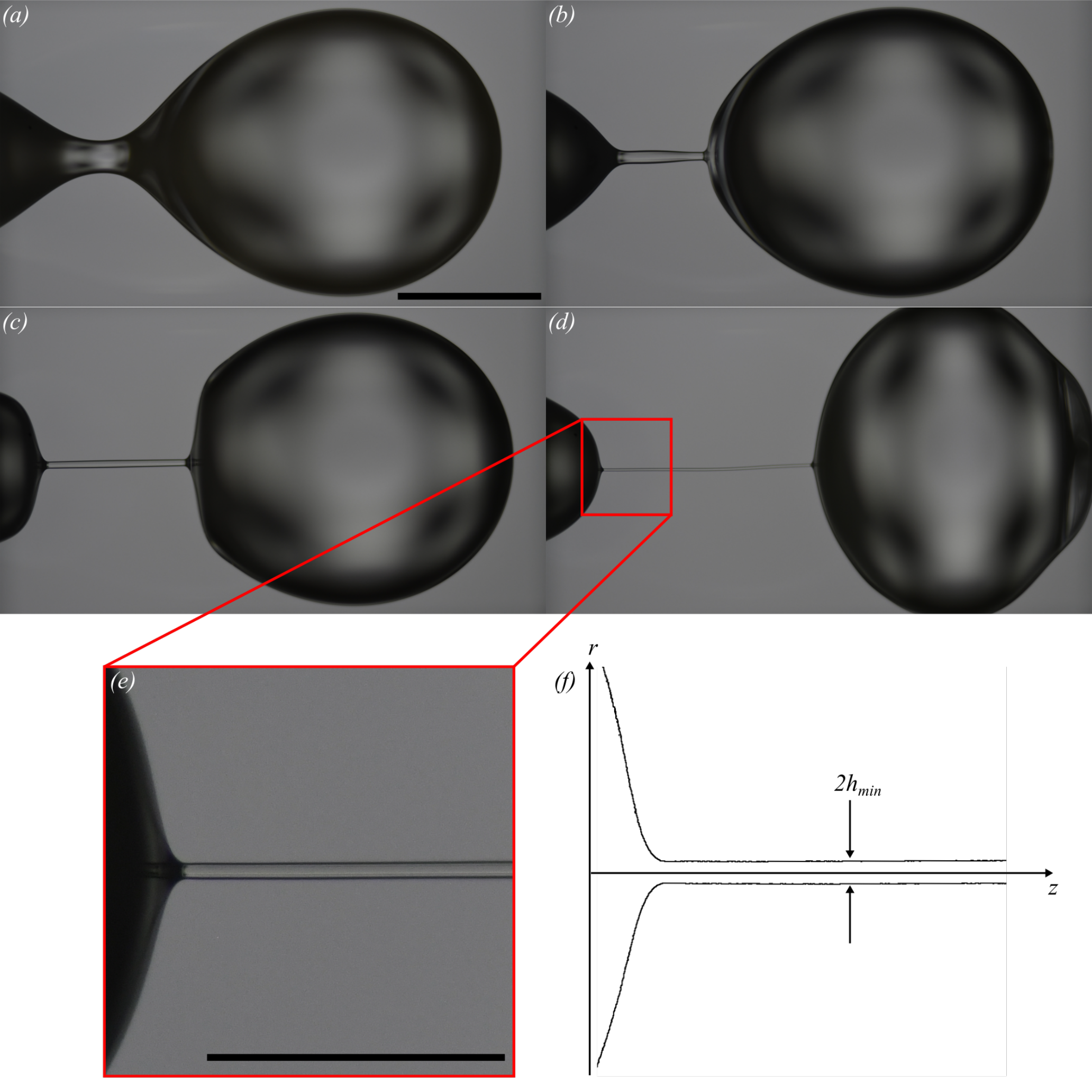}}
  \caption{High-resolution photographs of a pendant drop of PEO solution ($M_{w}$ = 4 $\times 10^{6} g/mol$ $C_{p}$= 10 wppm) breaking from a nozzle of diameter $h_{0}$ = 2 mm. Time between subsequent panels (a)-(d) is 1 ms. Scale bar is 2 mm. Panel (e) highlights the region of interest from which we extract the profile shown in (f). Scale bar is 1mm.}
\label{fig:PhotoBreakup}
\end{figure}

We used a Phantom V1 fast camera (frame rate 10,000 fps) to record the dynamics of the filament thinning at a high temporal resolution, and a full-frame camera ($8256 \times 4640~  \rm{pix}^{2}$, Nikon D850) equipped with a $5\times$ microscope lens to obtain a very good spatial resolution of the polymer interface during its detachment. In order to be able to capture high-quality picture of the interface during the fast thinning ($\sim$ms), we use a flash light (Vela one) with $1 \mu s$ flash duration. The camera and the flash light are coupled to a trigger to which we can control the delay to the next breakup event with a very good accuracy. The delay between two flashes is controlled through a precise delay line (Digital Delay Generator, DG535 Stanford Research Systems) that delays the initial trigger pulse (the event is triggered electronically on each bottom edge of droplets falling in repetition) and allows a delay resolution from 5ps to 1000s. The limiting factor here is the flash duration of 1 $\mu$s of the flashlight itself. The shutter of the camera is open for a `long' time during the flash, and so the read-out time of the camera is not a limiting step. This allow us to obtain a sequence of very highly resolved pictures (10 $\mu$m/pix). as one would obtain with a fast camera but with a much better resolution (see sequence of pictures in Fig.~\ref{fig:PhotoBreakup}). 

\begin{figure}
  \centerline{\includegraphics[scale=0.45]{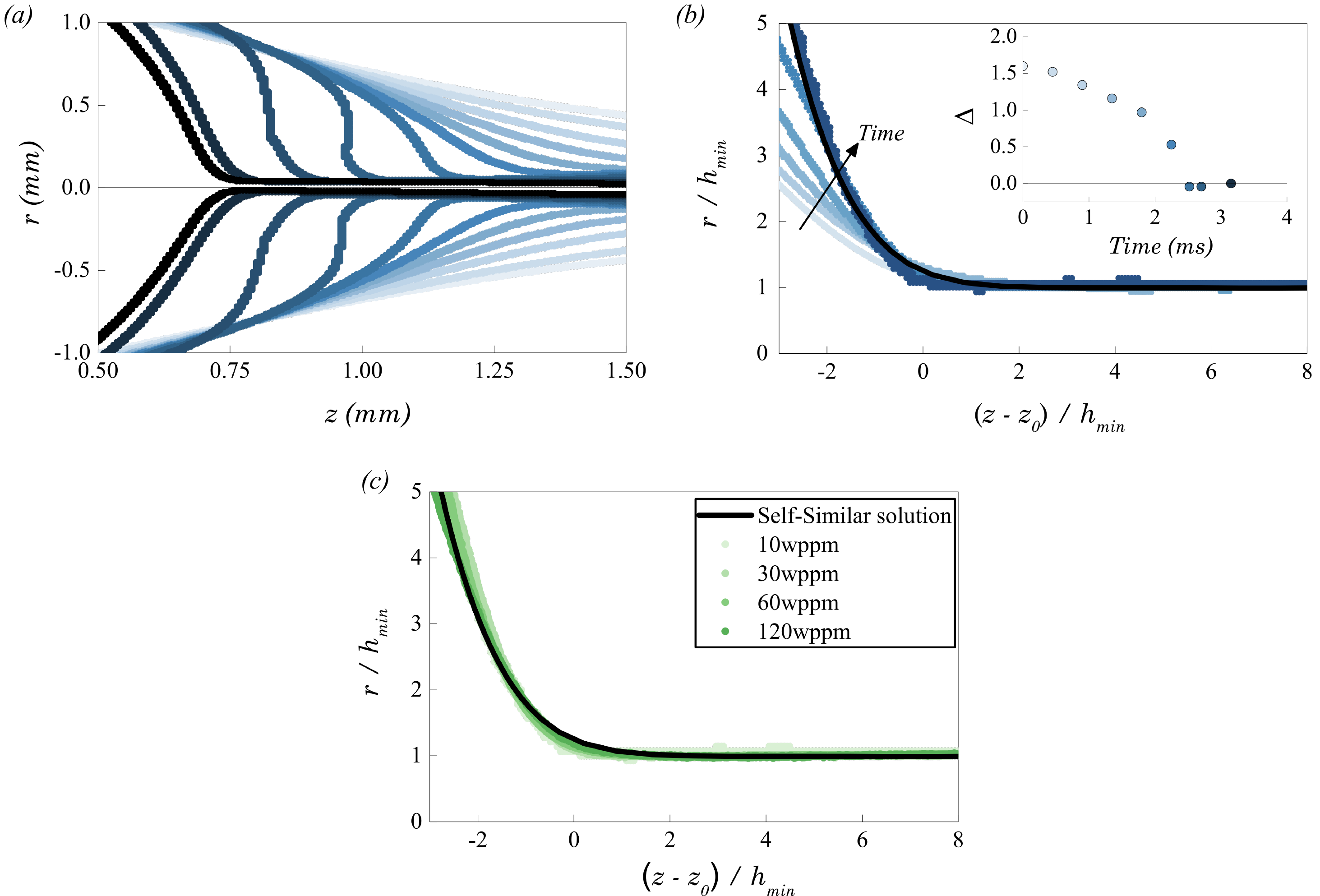}}
  \caption{(a) Time evolution of PEO filament profiles near the onset of the filament. Data are shown for subsequent times between each profile that are highlighted in figure \ref{fig:ThinningDynamics}(a) (star symbols) and for $C_{p}$ = 10 wppm. (b) Same profiles but rescaled by the minimum neck radius $h_{min}$ and with $z_{0}$ the location for which the experimental profiles beyond a time threshold collapse onto each other. The inset shows the convergence of the quantity $\Delta (t)$ toward the self-similar solution. (c) Post-threshold profiles for four polymer concentrations in the dilute regime. In (b) and (c), the solid black line indicates a universal self-similar solution calculated using the Oldroyd-B model.}
\label{fig:InterfaceProfile}
\end{figure}

Typical results are shown in Figures \ref{fig:ThinningDynamics} and \ref{fig:PhotoBreakup}. The thinning dynamics (Fig. \ref{fig:ThinningDynamics}) of 
PEO show an initial thinning similar to that of a low-viscosity Newtonian fluid. Subsequently, a very long and slender cylindrical filament is formed. In this elasto-capillary thinning regime, the dynamics slows down dramatically. Both the Oldroyd-B model and experiments show that in this regime the minimum neck radius $h_{min}$ as a function of time can be described as

\begin{equation}
h_{min} = h_{0} e^{-t/3 \lambda_{0}} ,
\label{eq:elastocapillaryregime}
\end{equation}

with $\lambda_{0}$ the longest relaxation time of the polymer solution \citep{Anna2001,Amarouchene2001}. In the range of (very) diluted concentrations studied here, $\lambda_{0}$ varies with concentration between $\sim$ 1 ms and 50 ms. Even though in the dilute limit a variation of the relaxation time is not predicted by theory, this is in fact commonly observed in experiments \citep{Clasen2006}. This is confirmed by plotting the dependence of the relaxation time $\lambda_{0}$ with the polymer concentration $C_{p}$ in figure~\ref{fig:ThinningDynamics}(b) show that it follows a power law dependence as reported by \cite{Clasen2006} for dilute polymer solutions. 

Figure~\ref{fig:PhotoBreakup} shows a typical sequence of pictures of a droplet's interface, obtained using the combination of fast camera and triggered flash to increase the resolution. We use these photographs to extract the interface profile with a homemade algorithm in Matlab; this procedure allows determining the edges of the polymer solution as shown in Figure~\ref{fig:PhotoBreakup}(f). 

\section{Results}

In Figure~\ref{fig:InterfaceProfile}(a) we show the evolution of the interface profile during filament thinning, near the drop that forms due to destabilization. Here, $r$ is the filament radius and $z$ the direction along the filament. In Figure~\ref{fig:InterfaceProfile}(b), we plot the same profiles but rescaled by the minimum neck radius $h_{min}$ and with $z_{0}$ an adjustable parameter that represents the location for which the experimental profiles collapse best. As time progresses, the interface shape converges to a universal shape. Comparing this shape to the recent viscoelastic calculations using the Oldroyd-B model (\cite{Eggers2020})., we find an excellent agreement: the profiles converge to the same universal self-similar solution profile indicated by the black line. We quantify the threshold to the self-similar solution by measuring the distance $\Delta (t)$ of the experimental profiles to the self-similar curve at $r/h_{min}=\exp(1)$. This quantity is shown in the inset of figure \ref{fig:InterfaceProfile}(b) and converges to a constant value at the moment where the elasto-capillary regime is reached. Since the onset is set by the elasto-capillary time, a dependence with the polymer concentration $C_{p}$ is expected. In Figure~\ref{fig:InterfaceProfile}(c), we show the profiles after converging for different polymer concentrations. They fall onto each other, confirming the self-similarity of the polymer thread interface. Here, again, agreement with theory (black line) is excellent; this also confirm that the profile grows exponentially as discussed in details in \cite{Eggers2020}.

Solving the constitutive equation for an Hookean dumbbell model, a quadratic dependence of the first normal stress difference $N_1$ on shear rate $\dot{\gamma}$ can be obtained \citep{Bird1987}:

\begin{equation}
N_{1}(\dot{\gamma}) = \psi_{1} \dot{\gamma}^{2} = 2 n k_{b} T \lambda_{0,\rm{rheo}}^{2} \dot{\gamma}^{2}~,
\label{eq:NormalStressOlroydB}
\end{equation}

\begin{figure}
  \centerline{\includegraphics[scale=0.95]{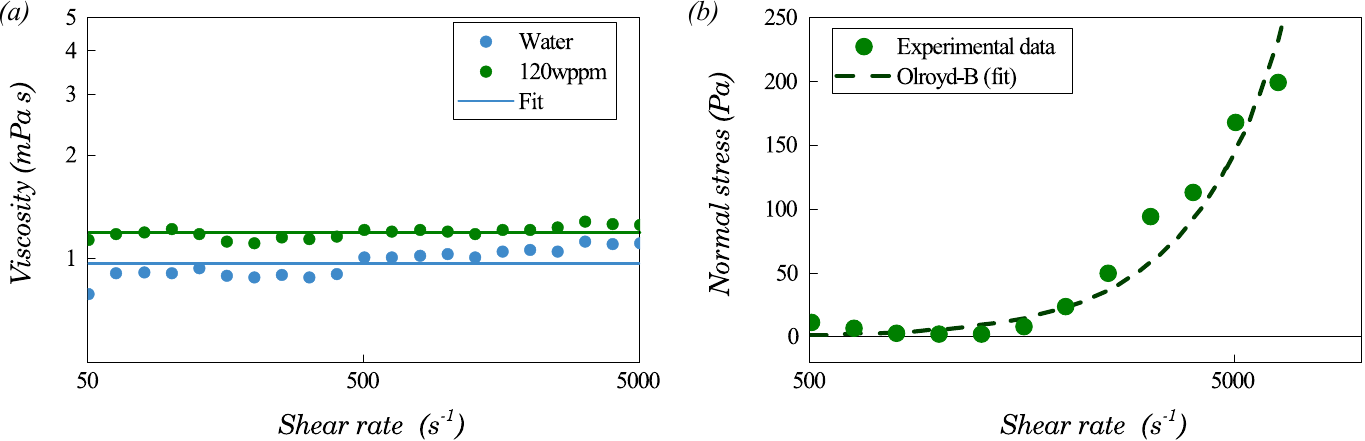}}
  \caption{(a) Shear rheology of a PEO solution (green symbols; $M_{w}=4 \times 10^{6}$ g mol$^{-1}$, $C_p$ = 120 wppm) compared to water (blue symbols) for shear rate values allowed by our experimental setup. (b) Magnitude of the first normal stress difference $N_{1}$ as a function of shear rate for a PEO solution ($C_p$ = 120 wppm). Dashed line is a fit of the Oldroyd-B model (Eq.~\ref{eq:NormalStressOlroydB}).}
\label{fig:Rheology}
\end{figure}

with $T$ denoting temperature, $\psi_1$ is the first normal stress coefficient, $n$ the number density of polymer molecules (Table \ref{tab:FluidParameters}) and $k_B$ Boltzmann's constant. The rheology measurements (Anton Paar, MCR 302) shown in Fig.~\ref{fig:Rheology}(a) show that the increase in polymer concentration slightly influence the shear viscosity, in agreement with Einstein expansion, whereas it significantly affects the normal force (Fig.~\ref{fig:Rheology}(b)), from which we can extract the relaxation time of the solution $\lambda_{0,\rm{rheo}}$ = 7.4 ms \citep{Lindner2003}. This is a factor $\sim$ 2 below the value obtained with the relaxation time obtained from the pinch-off experiment ($\lambda_{0}$ = 14.5 ms). Such discrepancies have been already been reported before by \cite{Clasen2006}. The primary reason for the discrepancy is that high molecular weight polymers invariably exhibit an significant polydispersity which may result in multiple timescales; the elongational flow is more sensitive to the longest time scale, whereas the shear flow probes an average time scale. We can nevertheless evaluate the ratio between these two time constants using the multimode Zimm model:

\begin{equation}
    \langle \lambda \rangle \approx \lambda_{0,\rm{rheo}} = \frac{1}{N} \sum_i \lambda_{i}= \frac{1}{N} \sum_i \frac{\lambda_{0}}{i^{2+\tilde{\sigma}}}
\end{equation}

where $N$ is the total number of modes and $\tilde{\sigma}$ is a measure of the hydrodynamics interaction and $\tilde{\sigma}\approx$ -0.4 in our dilute polymer solutions \citep{Anna2001}. This time quickly decays for the higher modes; evaluating the ratio between the two time scales for the five first modes gives a good approximation. In fact, this gives a value of $\lambda_{0,\rm{rheo}} / \lambda_{0}$ $\approx$ 3 which is a good estimate of what we find experimentally and which is a value also reported experimentally by \cite{Liang1994}.

\section{Conclusions}

In conclusion, we have studied the destabilisation of diluted polyethylene oxide (PEO) solution in water using a camera setup allowing us to visualize the droplet and filament shapes with very high resolution. We find that during destabilization of a polymer droplet initially attached to a capillary, the interface converges to a self-similar shape independent of time or polymer concentration, which agrees very well with the theoretical prediction of \cite{Eggers2020}, based on the Oldroyd-B model. This contrasts with an earlier discrepancy, observed in \cite{Turkoz2018}, between full numerical simulations of the Oldroy-B model and the earlier experiments of \cite{Clasen2006}. Three main differences between our experiment and that of \cite{Clasen2006} can been mentioned that might explain the observed discrepancy. (i) In the work presented here we are now able to work at significantly higher spatial and temporal resolution than they could do at this time. (ii) The second important difference to note is the solvent used that is much more viscous than the one we use in our study (water) that also exhibit non-Newtonian behavior `Boger fluid'. (iii) They used Caber device to impose an extensional deformation to their samples. For that purpose, they used endplates diameters up to 6mm (3 times our nozzle diameter $D_{min}$). As a consequence, gravitational and inertial effects may have an effect on the overall extensional flow: it consequently may generates additional flows during the thinning of the filament and affects the profile of the interface as discussed for instance in \cite{Brady1982}.

\end{document}